\documentclass{ws-mpla}

\begin{document}

\markboth{Yuri Shtanov and Alexander Viznyuk} {Mirror Branes}

%
\catchline{}{}{}{}{}
%

\title{MIRROR BRANES}

\author{\footnotesize YURI SHTANOV}

\address{Bogolyubov Institute for Theoretical Physics \\ Kiev, 03143, Ukraine \\
shtanov@bitp.kiev.ua}

\author{ALEXANDER VIZNYUK}

\address{Bogolyubov Institute for Theoretical Physics \\ Kiev, 03143, Ukraine \\
viznyuk@bitp.kiev.ua}

\maketitle

\pub{Received 3 September 2004}{Revised 5 June 2005}

\begin{abstract}
The Randall--Sundrum two-brane model admits the flat-brane Lorentz-invariant
vacuum solution only if the branes have exactly opposite tensions.  We pay
attention to this condition and propose a generalization of this model in which
two branes are described by actions of the same form and with the same matter
content but {\em with opposite signs}. In this way, the relation between their
tensions (which are their vacuum energy densities) is naturally accounted for.
We study a simple example of such a model in detail. It represents the
Randall--Sundrum model supplemented by the Einstein scalar-curvature terms in
the actions for the branes. We show that this model is tachyon-free for
sufficiently large negative values of the brane cosmological constant, that
gravitational forces on the branes are of opposite signs, and that physically
most reasonable model of this type is the one where the five-dimensional
gravity is localized around the visible brane.  The massive gravitational modes
in this model have ghost-like character, and we discuss the significance of
this fact for the quantum instability of the vacuum on the visible brane.

\end{abstract}

\ccode{PACS Nos.: 04.50.+h}

\def\lsim{\
  \lower-1.5pt\vbox{\hbox{\rlap{$<$}\lower5.3pt\vbox{\hbox{$\sim$}}}}\ }
\def\gsim{\
  \lower-1.5pt\vbox{\hbox{\rlap{$>$}\lower5.3pt\vbox{\hbox{$\sim$}}}}\ }

\section{Introduction:  the model}

The original Randall--Sundrum (RS) two-brane model\cite{RS1} describes vacuum
branes embedded in (or, equivalently, bounding) the five-dimensional AdS
spacetime and has the action
\begin{equation} \label{rs-action}
S_{\rm RS} = M^3 \!\! \int\limits_{\rm bulk}\!\! \left( {\cal R} - 2
\Lambda_{\rm b} \right) \ - \!\!\! \int\limits_{\rm brane1}  \!\!\! 2 \sigma \
+ \!\!\! \int\limits_{\rm brane2} \!\!\! 2 \sigma \, ,
\end{equation}
in which we omit the obvious integration volume elements and the boundary terms
with brane extrinsic curvature.  Here, $M$ denotes the bulk Planck mass,
$\Lambda_{\rm b}$ is the bulk cosmological constant, and the constant $\sigma$
is the vacuum energy density (brane tension) of one of the branes (brane1),
which is exactly opposite to the brane tension of the other brane (brane2). The
model admits the flat-brane Lorentz-invariant vacuum solution only under the
constraint\cite{RS1,RS2}
\begin{equation} \label{lambda-rs}
\Lambda_{\rm RS} \equiv {\Lambda_{\rm b} \over 2} + {\sigma^2 \over 3 M^6} = 0
\, ,
\end{equation}
which has to be postulated in the theory.\footnote{Flat-brane vacuum solutions
without Lorentz invariance of the bulk can be constructed without this
constraint.\cite{notune}}

The fact that two brane tensions are exactly equal by absolute value and
opposite in sign is also of crucial importance for the existence of flat-brane
solutions and requires explanation.   In this paper, we generalize the
Randall--Sundrum model (\ref{rs-action}) in such a way that two branes are
described by the same form of action with the same matter content but {\em with
the opposite signs}. In this way, the symmetry relation between their vacuum
energy densities is naturally accounted for.  Thus, we propose to consider the
generic action of the form
\begin{equation} \label{generic}
S =  \int\limits_{\rm bulk} {\cal L}_{\rm bulk} \left( g_{ab}, \Phi \right) \ +
\!\!\! \int\limits_{\rm brane1} \!\!\! {\cal L} \left( h_{ab}, \phi_1 \right) \
- \!\!\!\int\limits_{\rm brane2} \!\!\! {\cal L} \left( h_{ab}, \phi_2 \right)
\, ,
\end{equation}
where $\Phi$ denotes the fields in the bulk other than the metric field
$g_{ab}$. It is important to emphasize that the Lagrangians ${\cal L}$ have the
same form for two branes, with the field contents $\{\phi_1\}$ and $\{\phi_2\}$
being equivalent.  Therefore, their vacuum energy densities on the same metric
background will be exactly the same.

Were it not for the bulk part in action (\ref{generic}), we would have totally
decoupled actions for two non-interacting worlds. The relative signs between
the actions in this case would not matter at all.  The presence of the bulk
means a complicated interaction between the branes and also breaks the
equivalence symmetry between them since the physical constants in the bulk
Lagrangian are of specific signs.

The Lagrangian ${\cal L}$ in (\ref{generic}) is assumed to be written in the
``standard'' form.  In particular, the kinetic terms of the physical matter
fields enter it with the conventional signs.  The brane tension $\sigma$ is
then unambiguously defined as the vacuum energy density of the ``standard''
Lagrangian ${\cal L}$, i.e., on the background of the flat metric $h_{ab} =
\eta_{ab}$,
\begin{equation}
\left\langle {\cal L} \left(\eta_{ab}, \phi \right) \right\rangle_{\rm vac} = -
2 \sigma \, .
\end{equation}

A particular case of (\ref{generic}) that we are going to consider in this
paper is the generalization of the RS model (\ref{rs-action}) by adding the
matter Lagrangians and the Einstein scalar-curvature terms on the branes. In
this case, we have the action (again omitting the boundary terms with brane
extrinsic curvature)
\begin{equation} \label{action}
S =  M^3 \!\!\! \int\limits_{\rm bulk} \!\! \left( {\cal R} - 2 \Lambda_{\rm b}
\right) + \!\!\!\!\! \int\limits_{\rm brane1} \!\!\! \left[ \zeta R - 2 \sigma
+  L \left( h_{ab}, \phi_1 \right) \right] - \!\!\!\!\! \int\limits_{\rm
brane2} \!\!\! \left[ \zeta R - 2 \sigma + L \left( h_{ab}, \phi_2 \right)
\right] \, ,
\end{equation}
where we have explicitly introduced the vacuum energy density (brane tension)
$\sigma$ so that $\left\langle L \left(\eta_{ab}, \phi \right)
\right\rangle_{\rm vac} = 0$. The sign of the bulk Planck mass $M$ can be
chosen arbitrarily by suitably choosing the overall sign in action
(\ref{action}).  It is natural to choose  $M$ to be positive.  Then the brane
which appears with positive sign in action (\ref{action}) (which is brane1 in
our case) can be called {\em positive brane}, and the one appearing with
negative sign can be called {\em negative brane}. The scalar-curvature terms
for the two branes enter the total action (\ref{action}) with the same constant
$\zeta$ but with opposite signs, according to our general principle expressed
by action (\ref{generic}). This can also be quite naturally explained if these
terms are regarded as induced by the quantum corrections from the matter
Lagrangians $L(h_{ab}, \phi)$.  Since the matter actions on the branes have
opposite signs but equivalent matter content, the effective gravitational
actions that they induce on the respective branes are exactly of the same form
but of opposite signs.

Braneworld models with arbitrary relative signs of the brane gravitational
actions were recently under consideration in
Refs.~\refcite{CD,Padilla,Smolyakov}. In this paper, we propose to consider
perfect ``mirror'' symmetry between the actions for the branes, where both the
gravitational and matter actions are exactly of the same form but of opposite
sign.  We repeat, however, that symmetry between the branes is, in general,
broken by the presence of the bulk.

We allow for positive as well as negative gravitational coupling $\zeta$ in
(\ref{action}) (its sign is defined with respect to the ``standard'' matter
Lagrangian $L$). The brane tension $\sigma$ {\em a~priori\/} can also be of any
sign. First of all, we show that the brane cosmological constant $\lambda =
\sigma / \zeta$ in the model (\ref{action}) under consideration must be {\em
negative\/} and sufficiently large by absolute value in order that tachyonic
gravitational modes be absent in the theory. This will be the issue of the next
section, in which we also consider the ghost-like character of the massless
radion and massive gravitational modes. In Sec.~\ref{linear}, we study the
linearized gravity on the flat two-brane background.  We will see that the
gravitational forces exerted by matter have opposite signs on the two branes.
This selects the brane with attractive gravity as the ``visible'' one, which
turns out to be the negative brane (brane2) in the terminology introduced
above. We will also see that it is physically preferable for the model to have
positive gravitational constant $\zeta$, hence, negative tension $\sigma$.
Thus, the five-dimensional gravity is localized around the physical brane,
which makes it the ``Planck'' brane in the conventional terminology. In
Sec.~\ref{vacuum}, we discuss the quantum instability of the vacuum caused by
the ghost-like character of matter on the hidden brane and of the massive
gravitational modes. To resolve this problem, one can introduce an explicit
Lorentz-violating ultraviolet cutoff as discussed in Refs.~\refcite{CHT,CJM}.
In Sec.~\ref{poten}, we present approximate expressions for the gravitational
potential on the visible brane induced by a static source on the visible or
hidden brane. In Sec.~\ref{discuss}, we provide a general discussion of the
model.

\section{Tachyonic modes and ghosts} \label{tachyon}

Tachyonic modes were studied in our previous paper\cite{SV} in a model quite
similar to (\ref{action}), namely, where the matter and induced-gravity actions
for two branes were not necessarily the same (the constants $\zeta$ and the
Lagrangians $L$ could be different)  but entered the total action with the {\em
same\/} sign.\footnote{The peculiar relation between the brane tensions was
assumed to be the same as in the Randall--Sundrum model, since it is required
to have the flat-brane solutions.} The equations for the tachyonic modes in the
present case can be derived in exactly the same way as it was done in
Ref.~\refcite{SV} (see also Ref.~\refcite{CD}).

The background RS solution is described by the metric
\begin{equation}\label{metric}
ds^2 = dy^2 + a^2(y) \eta_{\alpha\beta} dx^\alpha dx^\beta \, , \quad a(y) =
\exp (- k y ) \, , \quad k = {|\sigma| \over 3 M^3} = \sqrt{-{\Lambda_{\rm b}
\over 6}} \, ,
\end{equation}
with the two branes located at $y = 0$ and $y = \rho > 0$, respectively. Thus,
throughout the paper, we have $k > 0$ by definition so that the left brane
(located at $y =0$) is the ``Planck'' brane (the one at which the massless
graviton is localized), while the right brane (located at $y = \rho$) is the
``TeV'' brane in the commonly used terminology. Then, the ``Planck'' brane is
positive (negative) if the brane tension $\sigma$ is positive (negative).
Proceeding to the Fourier analysis in terms of the Lorentzian momenta
$p^\alpha$ and being interested in the tachyonic case ($p^\alpha p_\alpha > 0$
for our signature convention), we introduce the dimensionless variables
\begin{equation}\label{variables}
\mu = - {\zeta \sigma \over 3 M^6} \, , \quad  \alpha = e^{k \rho} \, , \quad s
= {\sqrt{p^\alpha p_\alpha} \over k}
\end{equation}
and arrive at the following equation for the tachyonic modes (see
Ref.~\refcite{SV}):
\begin{equation}\label{E}
E(s) \equiv D_1(s) - \mu s \Bigl[ D(s) - \alpha D_*(s) \Bigr] - \alpha \left(
\mu s \right)^2 D_2 (s) = 0 \, ,
\end{equation}
where
\begin{equation} \begin{array}{l} \label{ds}
D_1(s) = I_1 (\alpha s) K_1 (s) - I_1 (s) K_1 (\alpha s)  \, ,  \\ D(s) = I_2
(s) K_1 (\alpha s) + I_1 (\alpha s) K_2 (s)  \, ,  \\ D_*(s) = I_1 (s) K_2
(\alpha s) + I_2 (\alpha s) K_1 (s) \, ,  \\ D_2(s) = I_2 (\alpha s) K_2 (s) -
I_2 (s) K_2 (\alpha s)  \, .
\end{array} \end{equation}

Since $\alpha > 1$ for $k > 0$, all four functions defined in (\ref{ds}) are
strictly positive for positive $s$.  Then Eq.~(\ref{E}) implies that tachyonic
modes are absent in the case $\mu = 0$ (which includes the RS model). Thus, we
need to consider the case $\mu \ne 0$. The asymptotic limits of the function
$E(s)$ in this case can easily be calculated:
\begin{equation}\label{limits}
\lim_{s \to 0} E(s) = (1 - 2 \mu) \sinh \rho \, , \quad \lim_{s \to \infty}
E(s) = - \infty \, , \quad \mu \ne 0 \, .
\end{equation}
This immediately implies the existence of tachyonic modes for nonzero $\mu <
1/2$.

Thus, for $\mu \ne 0$, one can expect tachyonic modes to be absent only if $\mu
\ge 1/2$. We show that this condition is also sufficient. Employing the method
used in Ref.~\refcite{SV}, we introduce the function
\begin{equation}
\bar E (s, t) = D_1(s) - t \Bigl[ D(s) - \alpha D_*(s) \Bigr] - \alpha t^2 D_2
(s) \, , \quad s , t > 0 \, ,
\end{equation}
so that $\bar E (s, \mu s) \equiv E(s)$ for positive $\mu$, which is now under
consideration. Then, solving the equation
\begin {equation}
\bar E (s, t) = 0
\end{equation}
with respect to $t$, we obtain the expression for the single positive root:
\begin{equation}
\bar t (s) = {\sqrt{ \Bigl[ D(s) - \alpha D_* (s) \Bigr]^2 + 4 \alpha D_1(s)
D_2 (s)} - \Bigl[ D(s) - \alpha D_* (s) \Bigr] \over 2 \alpha D_2 (s) } \, .
\end{equation}
It can be verified that this one-parameter function is convex upwards for all
values of the parameter $\alpha > 1$, and its asymptotic behavior for small and
large $s$ is given by
\begin{equation}\label{asympt}
\bar t (s) = {s \over 2} + o ( s )\, , \quad s \to 0 \, ; \qquad \lim_{s \to
\infty} \bar t (s) = 1 \, .
\end{equation}
Solving Eq.~(\ref{E}) is equivalent to solving the equation
\begin{equation}\label{bart}
\bar t (s) = \mu s \, ,
\end{equation}
which, in view of (\ref{asympt}), has a positive root for $s$ if and only if $0
< \mu < 1/2$.

From the asymptotic limits (\ref{asympt}) and from Eq.~(\ref{bart}) one can see
what happens to the tachyon mode as the value of $\mu$ approaches the
boundaries of the interval $\left(0, \frac12\right)$. In the limit $\mu \to 0$,
the mass of the tachyon goes to infinity and, in the limit $\mu \to 1/2$, it
goes to zero.  We see that the value $\mu = 0$ is an isolated point of the
theory, and one does not have a physically consistent model in its vicinity.

Thus, model (\ref{action}) has no tachyonic gravitational modes either if $\mu
= 0$ (which is an isolated point) or if $\mu \ge 1/2$.  However, as will be
shown shortly, the theory is singular at $\mu = 1/2$; hence, this value should
also be excluded. The physical condition $\mu > 1/2$ can be written as
\begin{equation}\label{notachyon}
\zeta \sigma < {} - \frac 32 M^6 \, .
\end{equation}
This constraint implies sufficiently large {\em negative\/} value of the brane
cosmological constant
\begin{equation} \label{coco}
\lambda \equiv {\sigma \over \zeta} < - {3 M^6 \over 2 \zeta^2} \, .
\end{equation}
Below, we restrict our investigation to this case.

The issue of ghosts in the gravitational sector of the complementary theory
with positive value of $M$ but arbitrary signs of the brane gravitational
constants was considered in Ref.~\refcite{Padilla}, and we apply the results
obtained therein to our case. First, we start with the radion. The radion
degree of freedom in our formalism is connected with the possibility of brane
bending in the bulk.  After identifying the relevant physical degrees of
freedom, one can obtain the effective action for the radion $\phi_{\rm r}$ in
our model using the results of Ref.~\refcite{Padilla}:
\begin{equation}
S_{\rm rad} = {3 M^3 \left(e^{2 k \rho} - 1 \right) \over k (1 - 2 \mu) } \int
\phi_{\rm r} \Box \phi_{\rm r} \, dx \, .
\end{equation}
One can see that the radion effective action is singular in the case $\mu =
1/2$, which is one of the reasons why this value should be
excluded,\footnote{The gravitational part of model (\ref{action}) with the
additional relation $\mu = 1/2$ was recently under consideration in
Ref.~\refcite{Smolyakov}.  It was observed that the braneworld theory becomes
degenerate in this case, which was also previously noted in Ref.~\refcite{SV}.}
and, under the condition of absence of tachyonic modes $\mu > 1/2$, the radion
field is a ghost from the viewpoint of the positive brane (brane1) and is not a
ghost from the viewpoint of the negative brane (brane2).

Now, following Refs.~\refcite{Padilla,SV}, one can show that the massive
gravitational modes in the theory under investigation have ghost-like nature.
For free metric perturbations described by the transverse traceless modes
$\gamma_{\alpha\beta} (x, y)$ in the form
\begin{equation} \label{induced}
h_{\alpha\beta} (x, y) =  e^{-2 k y} \Bigl[ \eta_{\alpha\beta} +
\gamma_{\alpha\beta} (x, y) \Bigr]\, ,
\end{equation}
expanding the perturbation in the physical Fourier modes $\psi(m,y)$ such that
$p_\alpha p^\alpha = - m^2$ with the corresponding boundary conditions,
\begin{equation} \label{Fourier}
\gamma_{\alpha\beta} (x, y) = \sum_m \chi_{\alpha\beta} (m, x) \psi (m, y) \, ,
\end{equation}
one arrives at the following quadratic effective action (cf.\@ with
Refs.~\refcite{Padilla,SV}):
\begin{equation} \label{eff-action}
S = \frac12 \sum_m C_m \int dx\, \chi^{\alpha\beta} (m, x) \left( \Box - m^2
\right) \chi_{\alpha\beta} (m, x) \, .
\end{equation}
For the massless mode ($m = 0$), we have $\psi(0, y) \equiv {\rm const}$, and
the constant $C_0$ is given by
\begin{equation} \label{c0}
C_0 = {M^3 (1 - 2 \mu) \over 2k} \left( 1 - e^{-2k\rho} \right) [\psi (0, 0)]^2
\, ,
\end{equation}
The case $\mu = 1/2$ was already excluded when considering the radion, and now
we see that it can also be excluded by the requirement of the nonzero norm of
the zero-mode graviton. Then, the norm of the zero-mode graviton is negative in
the region $\mu > 1/2$, where tachyonic modes are absent in the theory. Thus,
just as the massless radion, the massless graviton is a ghost from the
viewpoint of the positive brane, and is not a ghost from the viewpoint of the
negative brane. For the massive modes ($m \ne 0$), one obtains the expression
\begin{equation} \label{cm}
C_m = {M^3 \over m^2} \int_0^\rho dy e^{-4ky} [\psi' (m, y)]^2 \, , \quad m \ne
0 \, ,
\end{equation}
which means that the massive modes are ghosts from the viewpoint of the
negative brane.  It is worth noting that matter with ghost-like properties is
currently being employed in the literature, although in a somewhat different
context.\cite{ghost}

\section{Linearized gravity in the zero-mode approximation} \label{linear}

The zero-mode approximation developed in Ref.~\refcite{SV} can be directly
applied to the theory given by (\ref{action}).  Depending on the signs of the
gravitational constant $\zeta$ and brane tension $\sigma$ [which should be
opposite, as follows from (\ref{notachyon}) or (\ref{coco})], the components of
the Einstein tensor in the zero-mode approximation on the positive
($G^+_{\alpha\beta}$) and negative ($G^-_{\alpha\beta}$) brane in the
coordinates $x^\alpha$ are given by the following expressions:
\begin{eqnarray}\label{Einstein1}
G^\pm_{\alpha\beta} &=& 8 \pi G_{\rm N}  \left[ T^{-}_{\alpha\beta} - \frac13
e^{\pm 2 k \rho} \left( h^\pm_{\alpha\beta} - {\partial_\alpha
\partial_\beta \over \Box^\pm} \right) T^{-} \right] \nonumber \\ &-&  8 \pi
G_{\rm N} e^{- 2 k \rho}  \left[ T^{+}_{\alpha\beta} - \frac13 e^{\pm 2 k \rho}
\left( h^\pm_{\alpha\beta} - {\partial_\alpha \partial_\beta \over \Box^\pm}
\right) T^{+} \right]  + \ldots \, , \\ \nonumber &{}& \zeta > 0 \, , \quad
\sigma < 0 \, ,
\end{eqnarray}
\begin{eqnarray}\label{Einstein2}
G^\pm_{\alpha\beta} &=& 8 \pi G_{\rm N} e^{- 2 k \rho}  \left[
T^{-}_{\alpha\beta} - \frac13 e^{\mp 2 k \rho} \left( h^\pm_{\alpha\beta} -
{\partial_\alpha \partial_\beta \over \Box^\pm} \right) T^{-} \right] \nonumber
\\ &-&  8 \pi G_{\rm N} \left[ T^{+}_{\alpha\beta} - \frac13 e^{\mp 2 k
\rho} \left( h^\pm_{\alpha\beta} - {\partial_\alpha \partial_\beta \over
\Box^\pm} \right) T^{+} \right]  + \ldots \, , \\ &{}& \nonumber \zeta < 0 \, ,
\quad \sigma > 0 \, .
\end{eqnarray}
Here we used the notation
\begin{equation}\label{Newton}
8 \pi G_{\rm N} = {2 k \over M^3 (2 \mu - 1) \left(1 - e^{- 2 k \rho} \right) }
\, ;
\end{equation}
$T^\pm_{\alpha\beta}$, $h^\pm_{\alpha\beta}$, and $\Box^\pm$ are the
stress--energy tensors, flat induced metrics, and the corresponding
D'Alembertians on the positive (``$+$'') and negative (``$-$'') branes,
respectively, and $T^\pm$ are the traces of the corresponding stress--energy
tensors.  The dots in (\ref{Einstein1}) and (\ref{Einstein2}) denote the
higher-derivative terms.  The D'Alembertians in the denominators of
(\ref{Einstein1}) and (\ref{Einstein2}) reflect the presence of the radion
degree of freedom in the theory (see Ref.~\refcite{SV} for more details). The
gravitational constant (\ref{Newton}) is naturally inversely proportional to
the norm of the zero-mode graviton (\ref{c0}).

It can be seen from (\ref{Einstein1}) and (\ref{Einstein2}) that matter on the
negative brane gravitates attractively, while that on the positive brane
gravitates repulsively (in other terms, the low-mass gravitons behave as ghosts
with respect to matter on the positive brane). It can also be seen that the
general-relativistic part of gravity is stronger by a factor of $e^{2 k \rho}$
on the brane around which the five-dimensional gravity is localized, i.e., on
the ``Planck'' brane (it is positive or negative brane according to the sign of
$\sigma$), in agreement with the original result by Randall and
Sundrum.\cite{RS1} Thus, in all cases, it is physically reasonable to consider
the negative brane as ``visible'' (describing the observable world), and the
positive one as ``hidden''.

In the case of positive constant $\zeta$ and negative $\sigma$, the effect of
the hidden-brane matter is exponentially suppressed on the visible (negative)
brane as a function of the distance between the branes. Hence, the case of
negative $\sigma$ seems to be physically the most appealing.

\section{Quantum instability of the vacuum caused by ghosts} \label{vacuum}

The ghost-like character of the matter on the hidden brane and of the massive
gravitons in the model under consideration is something to be worried about and
requires further investigation.  The ghost nature of hidden matter is connected
with the fact that the matter Lagrangians enter action (\ref{generic}) with
opposite signs. In this respect, we can note that the gravitational influence
of the hidden matter in the model (\ref{action}) is exponentially suppressed by
the factor $e^{-2 k \rho}$, which significantly alleviates this problem.

The problem of the ghost nature of the massive gravitons is more serious and
can lead to quantum instability of the vacuum. To avoid too strong instability
of the vacuum caused by the ghosts, one can consider introducing an explicit
Lorentz-violating ultraviolet cutoff in the theory (see Ref.~\refcite{CHT,CJM}
for a recent discussion).

\begin{figure}
\center{\includegraphics[width=.3\textwidth]{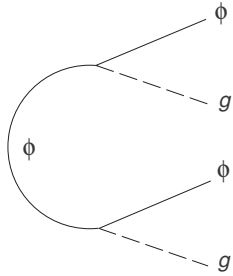}} \caption{Decay of
vacuum into two particles $\phi$ and two massive ghost gravitons $g$.}
\label{figure}
\end{figure}

As an example, in Fig.~\ref{figure}, we show a typical diagram of the process
in which a pair of matter particles $\phi$ are spontaneously created from the
vacuum on the visible (``Planck'') brane together with a pair of massive
gravitons $g$. The coupling of matter fields to massive gravitons in its
vertices is determined by the effective action (\ref{eff-action}) with the
constants $C_m$ given by (\ref{cm}).  Thus, each vertex of the process contains
the constant $\psi(m, 0) / \left| C_m \right|^{1/2}$, where $\psi (m, y)$ is
the wave function of the massive graviton in (\ref{Fourier}). Introducing also
the momentum ultraviolet cutoff $\Lambda$ similarly to Ref.~\refcite{CJM}, we
obtain the production rate for two graviton modes with absolute mass values
$m_1$ and $m_2$ on dimensional grounds,
\begin{equation}\label{gamma}
\displaystyle \Gamma_{m_1, m_2} \sim \left|{ \psi^2 (m_1, 0) \psi^2 (m_2, 0)
\over C_{m_1} C_{m_2}} \right| \Lambda^8  \, .
\end{equation}

The constant in (\ref{gamma}) can be estimated by using the asymptotic
expansions of the wave function in two mass regions as follows:
\begin{equation}\label{vertex}
\left| {\psi^2(m, 0) \over C_m} \right| \simeq {1 \over \mu^2 \alpha M^3}
\times \left\{
\begin{array}{ll}
k^3 / m^2 \, , \quad &m \gsim k \, , \medskip \\
m \, , &m \lsim k \, .
\end{array} \right.
\end{equation}

The interval between the modes is roughly equal to $\Delta m \simeq \pi k /
\alpha$ (see Ref.~\refcite{SV}); thus, summing expression (\ref{gamma}) over
the massive graviton modes, we estimate the total probability of particle
production as
\begin{equation}\label{gammatot}
\Gamma \sim {k^2 \Lambda^8 \over \mu^4 M^6 } \sim {\Lambda^8 \over (\zeta
\mu)^2} \sim {\Lambda^8 \over M_{\rm P}^4 \mu^2} \, ,
\end{equation}
where we have made the assumption that $(\mu - 1/2) \sim \mu$, i.e., that the
value of $\mu$ is not too close to $1/2$, and took into account that $\zeta =
M_{\rm P}^2$, where $M_{\rm P}$ is the effective Planck mass on the visible
brane.  Estimate (\ref{gammatot}) differs from that of Ref.~\refcite{CJM} by
the factor $\mu^2$ in the denominator.

The particles $\phi$ spontaneously produced from the vacuum in the process in
Fig.~\ref{figure} can, in particular, be photons.  Hence, by the same reasoning
as in Ref.~\refcite{CJM}, we can obtain the upper limit on the cutoff
$\Lambda$.  The spectral density of photons at $E \sim \Lambda$ created in the
process of Fig.~\ref{figure} is roughly given by the expression
\begin{equation}\label{density}
{d n \over d E} \sim {\Lambda^7 \over M_{\rm P}^4 \mu^2 } t_0 \, , \quad E
\lsim \Lambda \, ,
\end{equation}
where  $t_0 \sim H_0^{-1}$ is the age of the universe.  On the other hand, the
spectrum of photons is constrained by observations of the diffuse gamma ray
background by EGRET,\cite{egret} which has measured the differential photon
flux to be
\begin{equation}\label{flux}
{d F \over d E} = 7.3 \times 10^{-9} \left({E \over E_0} \right)^{- 2.1}
\left(\mbox{cm$^2$ s sr MeV} \right)^{-1}  \, ,
\end{equation}
where  $E_0 = 451$~MeV.  Demanding that (\ref{density}) does not exceed
(\ref{flux}) at $E \sim \Lambda$, we obtain the upper limit
\begin{equation}\label{limit}
\Lambda \lsim 3 \mu^{q}\, \mbox{MeV} \, , \quad q \approx \frac{2}{9.1} \approx
0.22 \, .
\end{equation}
This estimate differs from that of Ref.~\refcite{CJM} by the presence of the
factor $\mu^{q}$.  In principle, the value of $\mu$ is unbounded from above;
hence, the cutoff upper limit (\ref{limit}) can be made as large as possible.

Of course, these estimates have a preliminary character and should be supported
by the detailed analysis of the model.  Nevertheless, they give some hope that
the ghost character of the invisible matter and massive gravitational modes may
not be so serious in the model under consideration.

\section{Gravitational potential} \label{poten}

In this section, we present approximate expressions for the gravitational
potential on the visible brane induced by a static source on the visible or
hidden brane in the physically interesting case where the zero-mode graviton is
localized at the visible brane (positive $\zeta$ and negative $\sigma$), i.e.,
that the visible brane is the ``Planck'' brane. In doing this, we use the
results of our previous paper,\cite{SV} where the corresponding expressions are
derived for a generic two-brane model.

\subsection{Matter source on the visible brane}

In this case, we obtain the following gravitational potential $V(r)$ induced by
a static source of mass ${\cal M}$ on the visible brane:
\begin{equation}
V(r) = - {G {\cal M} \over r} \, , \quad G = G_{\rm N} \left(1 + {1 \over 3
\alpha^2 } \right) \, , \quad kr \gg \alpha \, ,
\end{equation}
\begin{equation}
V(r) = - {G {\cal M} \over r} \left( 1 - {2 \over 3 (2 \mu - 1) (kr)^2} \right)
\, , \ \ G = G_{\rm N} \left( 1 - {1 \over \alpha^2} \right) \, , \ \ 1 \ll kr
\ll \alpha \, ,
\end{equation}
where $G_{\rm N}$ is defined in (\ref{Newton}), and $\alpha$ in
(\ref{variables}). Note the correspondence with the structure of
(\ref{Einstein1}).

For $kr \ll 1$, we have the expression
\begin{equation} \label{newpot}
V (r) = - {G {\cal M} \over r}  - {k {\cal M} \over 3 \pi^2 \zeta} \left({15
\over 8} - \frac1\mu \right) \log \left[\left({15 \over 8} - \frac1\mu \right)
kr \right] \, , \ \ G = {(\mu - 2/3) \over 8 \pi \zeta (\mu -1/2)} \, ,
\end{equation}
where $\mu$ is defined in (\ref{variables}). As $\mu \to \infty$, we recover
the Newton's gravitational law with the standard four-dimensional gravitational
coupling $G = 1 / 8 \pi \zeta$.  Note that the Newtonian part of potential
(\ref{newpot}) is repulsive for $\mu$ in the narrow range $1/2 < \mu < 2/3$.
The logarithmic correction to this potential is only valid if the expression
$15/8 - 1 / \mu$ is not very small by absolute value.  The gravitational law on
small scales is four-dimensional [$V(r) \propto r^{-1}$] because the allowed
values of the brane gravitational constant $\zeta$ are sufficiently large, so
that $\mu > 1/2$.

\subsection{Matter source on the hidden brane}

In a similar way one can consider the case where the stationary matter with
mass ${\cal M}_*$ resides on the hidden brane. We obtain the following
expressions for the gravitational potential on the visible brane:
\begin{equation}
V(r) \approx  {4 G_{\rm N} {\cal M}_* \over 3 \alpha^2 r} \, , \quad kr \gg
\alpha \, ,
\end{equation}
\begin{equation}\begin{array}{r} \label{mu}
V (r) \approx  \displaystyle {2 k^2 {\cal M}_* \over 3 \pi^2 M^3 (2 \mu - 1)
\alpha^3 } \left[ \left( 1 + {c_1 \over \mu } - {c_2 \over \mu^2} \right) -
\left( c_3 + {c_4 \over \mu} - {c_5 \over \mu^2} \right) \left( { kr \over
\alpha } \right)^2 \right] \, , \medskip \\ kr \ll \alpha \, ,
\end{array}
\end{equation}
where the constants $c_n$ take the following approximate values:
\begin{equation}
c_1 \approx 1.3 \, , \quad c_2 \approx 0.35 \, , \quad c_3 \approx 0.06 \, ,
\quad c_4 \approx 1.12 \, , \quad c_5 \approx 0.24 \, .
\end{equation}
Note that the approximate potential is repulsive and that it does not have
Newtonian form for $kr \ll \alpha$.

\section{Discussion} \label{discuss}

The merit of generalization (\ref{generic}) proposed in this paper is that it
naturally takes into account the fact that two branes in the RS setup\cite{RS1}
have exactly opposite tensions.  We studied a particular version (\ref{action})
of the generic model (\ref{generic}) and have shown that it is tachyon-free if
the brane cosmological constant $\lambda = \sigma / \zeta$ is negative and
sufficiently large by absolute value, as given by Eq.~(\ref{coco}).  In this
case, the effective gravity is attractive on the negative brane (brane2) and is
repulsive on the positive brane (brane1).  Thus, a physically viable situation
obtains if the visible brane is taken to be the negative brane (brane2) in
action (\ref{action}).  If the gravitational constant $\zeta$ is positive
(hence, brane tension $\sigma$ is negative), then the five-dimensional gravity
is localized around the visible brane, where the effects of the hidden brane
are exponentially suppressed as functions of the distance $\rho$ between the
branes. The physical consequences of the repulsive gravity of matter of the
hidden brane remain to be clarified.

Reversing the overall sign of action (\ref{action}), one can see that the
preferable physical option from the viewpoint of the visible brane (brane2)
looks like that of negative bulk gravitational constant, which case was
discussed in our previous paper.\cite{SV} The negative sign of the bulk
gravitational constant with respect to the visible brane is motivated by some
braneworld models of dark energy, in particular, the braneworld model of {\em
disappearing dark energy\/}.\cite{SS} In this model, the expanding universe,
after the current period of acceleration, re-enters the matter-dominated regime
continuing indefinitely in the future. The model requires negative tension of
the visible brane, and the value of the bulk gravitational constant must then
be negative for the five-dimensional gravity to be localized around the brane.
Negative brane tension is required also for the existence of unusual
``quiescent'' singularities\cite{SS1} in the AdS-embedded braneworld models,
which occur during the universe expansion and are characterized by the {\em
finiteness\/} of the scale factor, Hubble parameter, and matter density.
Another instance of an interesting cosmological behavior in the case of
negative brane tension is the recently discussed ``cosmic mimicry'' in
braneworld models.\cite{mimicry} The present model of ``mirror branes'' has
also led us to the option of negative-tension visible brane around which the
five-dimensional gravity is localized.

The ghost-like character of the matter on the hidden brane and of the massive
gravitational modes in the model under consideration is something to be worried
about. The estimates obtained in Sec.~\ref{vacuum} show that the presence of
ghosts may not be a very serious problem. This issue, however, requires further
detailed investigation.

Other versions of the generic type (\ref{generic}) can be studied, among which
models with curvature terms on the branes and with the additional Gauss--Bonnet
term in the bulk,\cite{CD} models without the RS constraint
(\ref{lambda-rs}),\cite{Padilla} etc.  We mention them as a subject for future
investigations.

\end{document}